\documentclass[
    aps,
    fleqn,
    a4paper,
    superscriptaddress,
    preprint,
    preprintnumbers,
    showpacs,
    showkeys
]{revtex4}

\usepackage{graphicx}
\usepackage{picins}
\usepackage{amsmath}
\usepackage{amsfonts}
\usepackage[cp1250]{inputenc}

\begin{document}

\preprint{{\em Ferroelectrics\/}: Vol. 290, Pgs. 124-131 (2003) - {\em (Revised version)}}

\title{
Interrelation of domain wall contributions to dielectric, piezoelectric
and mechanical properties of a ferroic layer composite sample}

\author{A. \surname{Kopal}}
\email{antonin.kopal@tul.cz}
\author{P. \surname{Mokr\'{y}}}
\email{pavel.mokry@tul.cz}

\affiliation{Dept. of Physics, International Center for Piezoelectric Research, Technical University Liberec, Liberec 1, 461 17 Czech Republic}
\date{\today}

\begin{abstract}
Interrelations of domain wall (extrinsic) contributions to
dielectric, piezoelectric and mechanical properties of ferroic
samples are a hot subject of both theoretical and experimental
research. Recently,we have derived theoretical formulas for such
contributions, using model of composite layer sample: central
single-crystal ferroelectric-ferroelastic layer, isolated from
electrodes by passive layers. Here we present more general
results, discussing the dependence of the contributions on
geometric and material parameters of the composite and including
in a special case intrinsic piezoelectricity. We also discuss the
above-mentioned interrelation and compare our results with
measurements of all the contributions on the same RDP sample in a
wide temperature interval under the phase transition. It seems
that both calculated and observed results remind of the so called
Pippard-Janovec thermodynamic relations.
\end{abstract}


\maketitle

\section{Introduction}

The problem of extrinsic contributions to dielectric,
piezoelectric and mechanical properties has been addressed by many
authors. In prevailing number, only extrinsic permittivity has
been studied. Arlt et al. \cite{artA4:ref1,artA4:ref2} and
Herbiet {\em et al.} \cite{artA4:ref3} were the first, who
addressed the domain wall contributions to all involved
properties: permittivity $\varepsilon$, elastic compliance $s$ and
piezoelectric coefficient $d$ in piezoelectric ceramics (see also
a later work of Zhang {\em et al.} \cite{artA4:ref4}). In this
paper, we have in mind ferroelectric and ferroelastic crystals
with only two domain states, in particular, the KH$_2$PO$_4$
family. In our recent papers, we studied the equilibrium domain
structure in thin films \cite{artA4:ref5} and equilibrium
contributions to permittivity \cite{artA4:ref6}, piezoelectric
coefficient \cite{artA4:ref7} and to all these properties
including elastic compliance \cite{artA4:ref8}. In these papers
\cite{artA4:ref6,artA4:ref7,artA4:ref8} we have used the model
of passive layers (ferroic layer composite), theoretically
discussed first by Fedosov and Sidorkin \cite{artA4:ref9}, then
developed by Tagantsev {\em et al.} \cite{artA4:ref10} and
Bratkovsky and Levanyuk \cite{artA4:ref11,artA4:ref12}.

In this paper we shall concentrate on following questions: how the geometric
and material parameters of the composite effect the extrinsic contributions,
and what is the mutual relation of contributions to $\varepsilon$, $s$ and $d$
on the same sample.

In the following sections, we first introduce the model and basic
symbols for geometric and material properties of the sample. Then
we recapitulate briefly our method of theoretical derivation of
equilibrium extrinsic contributions, described in detail in
\cite{artA4:ref8}. Afterwards, we shall summarize and discuss
the new results, mentioned above and try to compare them with
experiment. Also, in a special case of equal material properties
of both passive and central layers, we include the effect of
intrinsic piezoelectricity. These three topics were not discussed
in \cite{artA4:ref8}.

\section{Model description}

%
\begin{figure}[t]
\begin{center}
\begin{minipage}{11cm}
  \includegraphics[width=85mm]{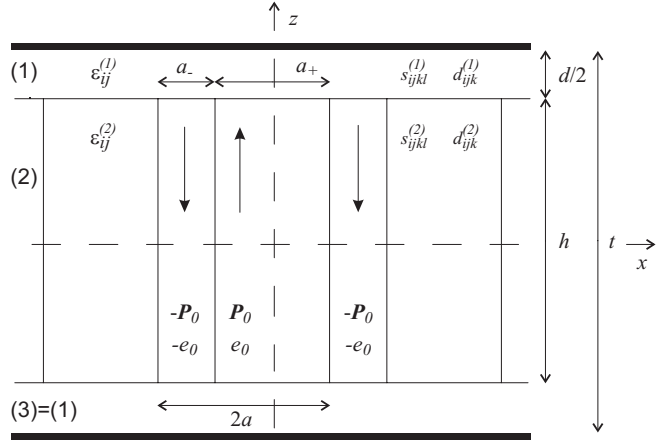}\vspace{3mm}
  \caption{Geometry of the model in $x-z$ plane. \label{artA4:fig01}}
\end{minipage}
\end{center}
\end{figure}
\begin{figure}[t]
\begin{center}
\begin{minipage}{11cm}
  \includegraphics[width=85mm]{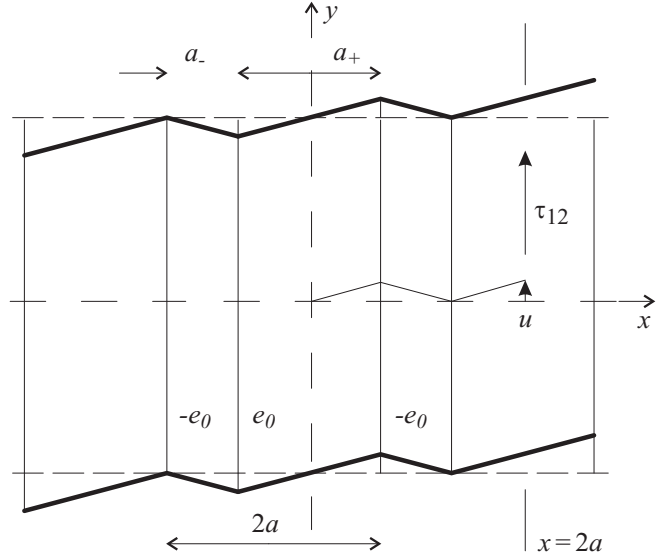}\vspace{3mm}
  \caption{Geometry of the model in $x-y$ plane. \label{artA4:fig02}}
\end{minipage}
\end{center}
\end{figure}

%
We consider a plate-like electroded sample of infinite area with
major surfaces perpendicular to the ferroelectric axis $z$.
Domains with antiparallel spontaneous polarization $P_{0,3} = \pm
P_0$ differ in the sign of spontaneous shear $e_{0,12} = \pm e_0$.
We suppose that usual linear state equations are valid, including
intrinsic piezoelectricity. For simplicity we consider the sample
to be elastically isotropic. The geometric parameters,
characterizing the sample and domain structure, are shown in Figs.
\ref{artA4:fig01} and \ref{artA4:fig02}.

%
%
We introduce the following symbols $\varepsilon^{(1)}_{33}$, the
component of relative permittivity of the passive layer;
$\varepsilon^{(2)}_{33}$, the same component for the central
ferroic layer; $s^{(1)}_{1212}$ and $s^{(2)}_{1212}$, the shear
component of the compliance tensor of the passive and central
ferroic layers, respectively; $d^{(1) }_{312}$ and
$d^{(2)}_{312}$, the piezoelectric coefficient of the passive and
central ferroic layers, respectively; $a_+$ resp. $a_-$, the width
of the corresponding domain; $2a$, the period of the domain
structure; $d$, the total width of the passive layers; $h$, the
width of the central layer; $u$, the shear displacement at $x =
2a$; $\tau = \tau_{12ext}$, the external shear stress acting on
the sample; $V$, the external voltage on the electrodes; $A = (a_+
- a_-)/(a_+ + a_-)$, the asymmetry factor of the domain structure,
expressing the displacement of the domain walls from the symmetric
position.

\section{Effective parameters of the sample}

In this short article, it is not possible to reproduce the whole
process of derivation of effective equilibrium permittivity
$\varepsilon^{\rm eff}_{33}$ at zero external stress, compliance
$s^{\rm eff}_{1212}$ at zero voltage $V$, and piezocoefficient
$d^{\rm eff}_{312}$ of the sample. The process is based on the
calculation of the total energy density $U$ of the sample, i.e.,
electrostatic energy, deformation energy and domain walls energy
per unite surface area of the sample in the $x-y$ plane, as a
function of free parameters $A$, $V$ and $u$ resp. $\tau$. In
equilibrium, the variation of the total energy of the isolated
system: {\em sample} + {\em electric source} + {\em mechanical
source} is zero. Solving this standard problem, we can find the
equilibrium reaction of the sample on external voltage or stress.
The process is described in \cite{artA4:ref8} and we shall
prepare more detailed publications in this direction.

Because of great complexity of the problem, we shall use the "thick sample"
approximation, valid when
\begin{equation}
    h,\, d \gg a.
    \label{artA4:eq01}
\end{equation}

In this case, we can neglect all higher Fourier components of the tensor of
deformation and electric field, except the constant one. Nevertheless, the
formulas for effective parameters are still very complex because of the
intrinsic piezoelectricity. If this is neglected, we get reasonably simple and
symmetric results:
\begin{eqnarray}
    \varepsilon^{\rm eff}_{33} &=&
        \varepsilon^{(1)}_{33}\,
        \frac{t}{d}\,
        \frac{
            P_0^2 \left(s^{(2)}_{1212}d + s^{(1)}_{1212}h\right) +
            e_0^2\varepsilon_0\varepsilon^{(2)}_{33}d
        }{
            e_0^2\varepsilon_0\, \left(\varepsilon^{(2)}_{33}d + \varepsilon^{(1)}_{33}h\right)
            + P_0^2 \left(s^{(2)}_{1212}d + s^{(1)}_{1212}h\right)
        },
    \label{artA4:eq02} \\
    s^{\rm eff}_{1212} &=&
        s^{(1)}_{1212}\,
        \frac{t}{d}\,
        \frac{
            e_0^2\varepsilon_0 \left(\varepsilon^{(2)}_{33}d + \varepsilon^{(1)}_{33}h\right) +
            P_0^2s^{(2)}_{1212}d
        }{
            e_0^2\varepsilon_0\, \left(\varepsilon^{(2)}_{33}d + \varepsilon^{(1)}_{33}h\right)
            + P_0^2 \left(s^{(2)}_{1212}d + s^{(1)}_{1212}h\right)
        },
    \label{artA4:eq03} \\
    d^{\rm eff}_{312} &=&
        s^{(1)}_{1212}\,
        \varepsilon^{(1)}_{33}\,
        \frac{h}{d}\,
        \frac{
            e_0P_0\varepsilon_0\, t
        }{
            e_0^2\varepsilon_0\, \left(\varepsilon^{(2)}_{33}d + \varepsilon^{(1)}_{33}h\right)
            + P_0^2 \left(s^{(2)}_{1212}d + s^{(1)}_{1212}h\right)
        }.
    \label{artA4:eq04}
\end{eqnarray}

For thin films, when condition (\ref{artA4:eq01}) does not hold,
numerical methods should be applied.

In the homogeneous composite case $\varepsilon^{(1)}_{33} =
\varepsilon^{(2)}_{33} = \varepsilon_{33}$, $s^{(1)}_{1212} =
s^{(2)}_{1212} = s_{1212}$, $d^{(1)}_{312} = d^{(2)}_{312} =
d_{312}$, we need not to neglect intrinsic piezoelectricity and
results are still very simple and symmetric:

\begin{eqnarray}
    \varepsilon_0\varepsilon^{\rm eff}_{33} &=&
        \varepsilon_0\varepsilon_{33}
        +
        \frac{h}{d}\,
        \frac{
            P_0^2
        }{
            P_0^2/(\varepsilon_0\varepsilon_{33}) + e_0^2/s_{1212} + P_0e_0/d_{312}
        },
    \label{artA4:eq05} \\
    s^{\rm eff}_{1212} &=&
        s_{1212}
        +
        \frac{h}{d}\,
        \frac{
            e_0^2
        }{
            P_0^2/(\varepsilon_0\varepsilon_{33}) + e_0^2/s_{1212} + P_0e_0/d_{312}
        },
    \label{artA4:eq06} \\
    d^{\rm eff}_{312} &=&
        d_{312}
        +
        \frac{h}{d}\,
        \frac{
            e_0P_0
        }{
            P_0^2/(\varepsilon_0\varepsilon_{33}) + e_0^2/s_{1212} + P_0e_0/d_{312}
        }.
    \label{artA4:eq07}
\end{eqnarray}

\section{Discussion}

Equations (\ref{artA4:eq02})-(\ref{artA4:eq04}) express the
dependence of sample effective parameters on both material and
geometric parameters of the layer composite. At first, we can
mention that for homogeneous composite, these relations transform
to (\ref{artA4:eq05})-(\ref{artA4:eq07}), without the terms
containing $d_{312}$. Intrinsic piezoelectricity is neglected in
equations (\ref{artA4:eq02})-(\ref{artA4:eq04}). Second, we can
find the effective parameters of the ferroic sample with thick
metal electrodes and without dielectric passive layers, taking the
limit of infinite $\varepsilon^{(1)}_{33}$ in
(\ref{artA4:eq02})-(\ref{artA4:eq04}). Third, the denominator on
the right in discussed relations is the same and we can easily
express the ratio of effective parameters.

Now, we can explore the limits $e_0\rightarrow 0$ (pure ferroelectric) or
$P_0\rightarrow 0$ (pure ferroelastic). In the first case we get
\begin{eqnarray}
    \varepsilon^{\rm eff}_{33} &=&
        \varepsilon^{(1)}_{33}\,t/d,
    \label{artA4:eq08} \\
    s^{\rm eff}_{1212} &=&
        \frac{s^{(1)}_{1212}s^{(2)}_{1212}\,t}{s^{(2)}_{1212}d + s^{(1)}_{1212}h}.
    \label{artA4:eq09}
\end{eqnarray}

In equation (\ref{artA4:eq08}), we find the simple result that
domainwalls motion is nownot damped by mechanical interaction with
passive layers, as reported in \cite{artA4:ref6}. In Eq.
(\ref{artA4:eq09})
 we can easily recognize the pure intrinsic compliance of the composite,
without any extrinsic contribution. In the second case we get
\begin{eqnarray}
    \varepsilon^{\rm eff}_{33} &=&
        \frac{\varepsilon^{(1)}_{33}\,\varepsilon^{(2)}_{33}\, t}{\varepsilon^{(2)}_{33}d + \varepsilon^{(1)}_{33}h}.
    \label{artA4:eq10} \\
    s^{\rm eff}_{1212} &=&
        s^{(1)}_{1212}\, t/d.
    \label{artA4:eq11}
\end{eqnarray}

Now we have got an inverse situation: in equation
(\ref{artA4:eq10}), there is pure intrinsic permittivity, without
any extrinsic contribution, while equation (\ref{artA4:eq11}) is a
simple formula that can be implicitly found in
\cite{artA4:ref8}. Naturally, $d^{\rm eff}_{312} = 0$ in both
limits. We should mention, that simple limit $d\rightarrow 0$ in
equations (\ref{artA4:eq08}) and (\ref{artA4:eq11}) does not work,
because condition (\ref{artA4:eq01}) for the thick sample
approximation is violated. In the general case, $P_0 \neq 0$, $e_0
\neq 0$, the effective permittivity and compliance take the value
in the corresponding intervals (\ref{artA4:eq10}) to
(\ref{artA4:eq08}), or (\ref{artA4:eq09}) to (\ref{artA4:eq11}).

Let us now turn to the homogeneous case of equations
(\ref{artA4:eq05})-(\ref{artA4:eq07}), where intrinsic
piezoelectricity is included. First, we can see that intrinsic and
extrinsic contributions are now explicitly separated.We denote the
extrinsic contributions $\Delta\varepsilon_0\varepsilon$, $\Delta
s$, $\Delta d$. The interrelation of the contributions is now
rather simple:
\begin{equation}
    \Delta\varepsilon_0\varepsilon : \Delta s : \Delta d =
    P^2_0 : e^2_0 : e_0P_0.
    \label{artA4:eq12}
\end{equation}
From equation (12), we get the simple condition
\begin{equation}
    \frac{\Delta\varepsilon_0\varepsilon \Delta s}{\left(\Delta d\right)^2} = 1.
    \label{artA4:eq13}
\end{equation}

The authors of Ref. \cite{artA4:ref13} have measured all three
effective parameters on the same $RbH_2PO_4$ sample in the
temperature interval of 90 to 140 K under the critical point,
resulting in an experimental value of 1,05 for the ratio in
equation (\ref{artA4:eq13}). They have come to the similar result
on the base of much more elementary considerations. In
\cite{artA4:ref8}, we presented discussion of other facts
concerning the agreement of theory and that experiment.

Under the critical temperature of the ferroelectric phase transition, $P_0$ and
$e_0$ often undergo substantial changes with changing temperature. Nevertheless,
 their ratio is almost constant in a wide temperature interval. This can be
explained by simple physical considerations. The crystals of KDP family undergo
the first order phase transition, which is close to the second order one,
because discontinuous change of $P_0$ and $e_0$ is relatively small. From the
Landau-Ginzburg theory that is in a good agreement with experiment in this case,
 we get that both $P_0$ and $e_0$ are proportional to $(T_c - T)^{1/2}$ in the
region of temperature $T$ under the critical one $T_c$. As a
result, the ratios in equation (\ref{artA4:eq12}) should remain
also constant in this temperature interval. This is fully
confirmed by the above-mentioned measurements in Ref.
\cite{artA4:ref13}. The situation here is very similar to that
discussed in connection with the so-called Pippard-Janovec
relations (see the excellent theoretical work of Janovec (1966)
\cite{artA4:ref14}). It is obvious that the results, expressed
by equations (\ref{artA4:eq12}), (\ref{artA4:eq13}), are general
and independent on the model. On the other side, results
(\ref{artA4:eq02})-(\ref{artA4:eq07}) are partially characteristic
for our model because of the presence of geometric factors $a$ and
$b$ and the material factors, characteristic of the passive
layers.

\acknowledgements 

The authors are indebted to Prof. J. Fousek for valuable
information and cooperation and to Prof. V. Janovec for
illuminating discussions. This work was supported by the Grant
Agency of the Czech Republic, project GA\v{C}R 202/00/1245 and by the
Ministry of Education of Czech Republic, project MSM 242 200 002.


\begin{thebibliography}{99}

\bibitem{artA4:ref1} G. Arlt and N. A. Pertsev, {\em J. Appl. Phys.} {\bf 70}, 2283 (1991).
\bibitem{artA4:ref2} G. Arlt and H. Dederichs, {\em Ferroelectrics} {\bf 29}, 47 (1980).
\bibitem{artA4:ref3} R. Herbiet, U. Robels, H. Dederichs, and G. Arlt, {\em Ferroelectrics} {\bf 98}, 107 (1989).
\bibitem{artA4:ref4} Q. M. Zhang, H. Wang, N. Kim, and L. E. Cross, {\em J. Appl. Phys.} {\bf 75}, 459 (1994).
\bibitem{artA4:ref5} A. Kopal, T. Bahn\'{\i}k, and J. Fousek, {\em Ferroelectrics} {\bf 202}, 267 (1997).
\bibitem{artA4:ref6} A. Kopal, P. Mokr\'{y}, J. Fousek, and T. Bahn\'{\i}k, {\em Ferroelectrics} {\bf 223}, 127 (1999).
\bibitem{artA4:ref7} A. Kopal, P. Mokr\'{y}, J. Fousek, and T. Bahn\'{\i}k, {\em Ferroelectrics} {\bf 238}, 203 (2000).
\bibitem{artA4:ref8} P. Mokr\'{y}, A. Kopal, and J. Fousek, {\em Ferroelectrics} {\bf 257}, 211 (2001).
\bibitem{artA4:ref9} V. N. Fedosov and A. S. Sidorkin, {\em Sov. Phys. Solid State} {\bf 18}, 964 (1976).
\bibitem{artA4:ref10} A. K. Tagantsev, C. Pawlaczyk, K. Brooks, and N. Setter, {\em Integrated Ferroelectrics} {\bf 4}, (1994).
\bibitem{artA4:ref11} A. M. Bratkovsky and A. P. Levanyuk, {\em Phys. Rev. Lett.} {\bf 84}, 3177 (2000).
\bibitem{artA4:ref12} A. M. Bratkovsky and A. P. Levanyuk, {\em Phys. Rev. Lett.} {\bf 85}, 4614 (2000).
\bibitem{artA4:ref13} M. \v{S}tula, J. Fousek, H. Kabelka, M. Fally, and H. Warhanek, {\em J. Kor. Phys. Soc.} {\bf 32}, 758 (1998).
\bibitem{artA4:ref14} V. Janovec, {\em J. Chem. Phys.} {\bf 45}, 1874 (1966).

\end{thebibliography}
\end{document}